\documentclass[aps,prd,preprint,groupedaddress,showpacs]{revtex4}
\usepackage{amsmath}
\usepackage{amssymb}
\usepackage{graphicx}
\usepackage[utf8]{inputenc}
\usepackage[T1]{fontenc}
\usepackage{units}
\usepackage{textcomp}
\usepackage{graphicx}

\begin{document}

\title{Two-loop effective potential for the Wess-Zumino model in 2+1 dimensions}

\author{R. V. Maluf}
\email{rmaluf@fma.if.usp.br}
\affiliation{Instituto de F\'isica, Universidade de S\~ao Paulo,
Caixa Postal 66318, 05315-970, S\~ao Paulo, SP, Brazil}
\affiliation{Departamento de F\'isica - Universidade Federal do Cear\'a
C.P. 6030, 60455-760, Fortaleza, Cear\'a, Brazil}


\author{ A. J. da Silva}
\email{ajsilva@fma.if.usp.br }
\affiliation{Instituto de F\'isica, Universidade de S\~ao Paulo,
Caixa Postal 66318, 05315-970, S\~ao Paulo, SP, Brazil}


\begin{abstract}
By using superfield techniques, the effective potential of the $\mathcal{N}=1$ 
Wess-Zumino model in 2+1 dimensions is computed off-shell
up to two loops. It is shown that supersymmetry is not dynamically broken,
and that dynamical generation of mass does not occur
perturbatively. We also investigate the renormalization of the effective
potential and determine the renormalization
group gamma and beta functions, showing that this model is infrared free. 
Comparison with some other results in the literature is provided.
\end{abstract}


\maketitle


\section{\label{sec:intro}Introduction}
Although supersymmetry (SUSY) is a key concept in the physics of elementary particles and fields, it is not supported (up to now) by experimental evidences. So, any realistic model involving SUSY must include some mechanism of breakdown. Many different mechanisms of breakdown have been considered in the literature. For instance, the Minimal Supersymmetric Standard Model (MSSM) with explicit soft SUSY breaking operators has been suggested as a way of solving the scale of grand unification and the hierarchy problems \cite{key-1}. The breakdown due to instanton solutions \cite{key-2}, its connection with $R$-symmetry
breaking \cite{key-3} and with the Witten index \cite{key-4,key-5}
have also been intensely investigated along the years. Yet, several variations
or extensions of the models of O'Raifeartaigh and of Fayet-Iliopoulos
\cite{key-6}, which present spontaneous SUSY breaking, have been
considered and more recently, theories which exhibit metastable vacua with broken SUSY \cite{key-8} have
also been proposed. Another interesting question is whether a purely
perturbative mechanism, i.e., a dynamical symmetry breaking induced
by radiative corrections can be achieved (in this case, a mass scale would be dynamically generated).

In $3+1$ spacetime dimensions this possibility is ruled out by nonrenormalization theorems
\cite{key-10}. On the other hand, in $2+1$ dimensions such restriction
(at least for $\mathcal{N}=1$ SUSY) does not exist \cite{key-11, key-11.2}. The usual way of investigating the vacuum structure
in quantum field theory involves the calculation of the effective
potential \cite{key-12}. Recently the two-loop effective potential for the
three-dimensional $\mathcal{N}=2$ Wess–Zumino (WZ) model was evaluated in Ref. \cite{key-12.1}. For the case $\mathcal{N}=1$, the effective potential for the WZ model and massless electrodynamics up to one-loop were first calculated in \cite{key-13} long ago. In both models, that author showed that neither SUSY nor the gauge invariance
are broken by radiative corrections up to one loop order. Nevertheless, in
$2+1$ dimensions, terms involving logarithms of the
classical fields only appear in two or more loops. Since these logarithmic
contributions have a crucial role in the dynamical symmetry breakdown,
the calculations must be carried up at least to two loops.

In the component field formalism the two-loop effective potential
of the WZ model was evaluated off-shell and on-shell in Refs.
\cite{key-14} and \cite{key-15}, respectively. In Ref. \cite{key-14},
it is reported a problem with the renormalization of the effective
potential: a divergent term which can not be absorbed
by rescaling of the classical Lagrangian appears. On the other hand, in Ref.
\cite{key-15}, difficulties with the renormalization are not found,
but it is claimed that SUSY is broken and a dynamical mass generation
takes place. In that paper, however, the evaluation of the
effective potential did not take into account radiative corrections
to the equation of motion of the auxiliary field \cite{key-16}. These
facts lead us to conclude that the renormalization and the vacuum
structure to the three-dimensional WZ model are issues not
yet satisfactorily answered.

The present work aims to calculate the two-loop effective potential
of the WZ model by using the superfield formulation. We claim that the 
renormalization of the effective potential with dimensional reduction regularization 
is achieved in the usual way. Moreover, we show
that SUSY is not broken and dynamical generation of mass
is not perturbatively consistent. 
We have also determined the beta function associated
with the fourfold self-interaction and verified that it agrees with the
result in Ref. \cite{key-17}, which is obtained
by direct calculation of the one-particle irreducible (1PI) Green's
functions in components fields. The anomalous dimension of the superfield is also determined.

The paper is organized as follows. In Sec. \ref{sec:TheModel}, the model is defined and the tree level potential is analyzed for different set-ups of the coupling constants. In Sec. \ref{sec:FeynmanRules}, the effective potential in one and two-loop order is calculated and its renormalization is analyzed, for the most general WZ model of a single real scalar superfield. In Sec. \ref{sec:4} the possibility of dynamical symmetry breakdown, for the (sub) model that is classically scale invariant, is studied with the conclusion that the symmetries are preserved. The beta function of the coupling constant is also calculated, showing that the model has a Landau pole in the UV limit. In Sec. \ref{sec:conclusions}, we summarize our conclusions. In  Appendix \ref{sec:Appendix-A:-TheZetaFuction} the $\zeta$-function method for the calculation of the one-loop contribution is outlined and in Appendix \ref{sec:Appendix-B:-Two-loopDiagrams} some details of the two-loops calculations are presented.

\section{\label{sec:TheModel}The model}

The most general renormalizable action for the $\mathcal{N}=1$ WZ model, containing a single real scalar superfield in $2+1$ dimensions is given by
\begin{equation}
{\cal S} [\Phi]=\int d^{5}z\left\{ -\frac{1}{4}D^{\alpha}\Phi D_{\alpha}\Phi+
W(\Phi)+\mathcal{L}_{CT}\right\} ,\label{eq1}
\end{equation}
where $W=a\Phi+\frac{1}{2}m\Phi^{2}+\frac{\lambda}{3!}\Phi^{3}+\frac{g}{4!}\Phi^{4}$ is the superpotential, $\Phi(x,\theta)=\phi(x)+\theta^{\alpha}\psi_{\alpha}(x)-F(x)\theta^{2}$ is a scalar superfield, 
$d^{5}z \equiv d^{3}xd^{2}\theta$ is the superspace element of volume
and $\mathcal{L}_{CT}$ is the counterterm Lagrangian.
Our conventions and notations for the superfield formalism are the
same as in Ref. \cite{key-18}. The mass dimensions of the scalar superfield and the coupling constants
are: $\left[\Phi\right]=\nicefrac{1}{2}$, $\left[\lambda\right]=\nicefrac{1}{2}$,
$\left[g\right]=0$, $\left[a\right]=\nicefrac{3}{2}$. When $\lambda=a=0$, the classical
action is invariant under the discrete symmetry transformation $\Phi\rightarrow-\Phi$ and if in addition $m=0$ the model is also classically scale invariant.

The component form of Eq. (\ref{eq1}) is easily obtained by doing the {\small $\theta$-integration: 
\begin{eqnarray}
{\cal S} & = &\int d^{3}x\left\{ \frac{1}{2}(\phi\square\phi+\psi^{\alpha}i\partial_{\alpha}^{\ \beta}\psi_{\beta}+F^{2})+m(\psi^{2}+\phi F)\right.\nonumber \\
 & +& \left.\lambda(\phi\psi^{2}+\frac{1}{2}\phi^{2}F)+\frac{g}{6}\phi^{3}F+\frac{g}{2}\phi^{2}\psi^{2}+aF+\mathcal{L}_{CT}\right\} .\label{eq:2}
\end{eqnarray}

The above action is invariant under the supersymmetry transformations
\begin{eqnarray}
\delta\phi & = &-\epsilon^{\alpha}\psi_{\alpha},\nonumber \\
\delta\psi_{\alpha} & = &-\epsilon^{\beta}(C_{\alpha\beta}F+i\partial_{\alpha\beta}\phi),\nonumber \\
\delta F & = &-\epsilon^{\alpha}i\partial_{\alpha}^{\ \beta}\psi_{\beta},\label{eq:2.1}
\end{eqnarray} where $\epsilon^{\alpha}$ is a constant fermionic parameter.

The tree level effective potential, as can be read directly from Eq. (\ref{eq:2}), is given by
\begin{equation}
V^{(0)}(\phi,F)=-\frac{1}{2}F^{2}-F S(\phi),\label{eq:3}
\end{equation}
where $S(\phi) \equiv W'(\phi)=(a+m\phi+\frac{\lambda}{2}\phi^{2}+\frac{g}{6}\phi^{3})$.
By eliminating the auxiliary field $F$ through its algebraic equation of motion
$0=\partial V^{(0)}/\partial F=-F-S(\phi)$, the classical potential becomes
only a function of the physical field $\phi$, such that
\begin{equation}
V^{(0)}(\phi)=\frac{1}{2}(S(\phi))^{2}  \geq 0.\label{eq:4}
\end{equation} 

As is well known, for any unbroken supersymmetric theory the vacuum
state must corresponds to a global minimum of the effective potential
with $S(\phi_{min})=0$ and $V(\phi_{min})=0$ \cite{key-19}.
For $g\neq0$ the model (at tree level) has a SUSY preserving phase,
since $S(\phi_{min})=0$ always has at least one real solution for
$\phi_{min}$. In this case, if $a=\lambda=0$ we have a minimum at
$\phi_{min}=0$ and if besides that, we also have $-6m/g>0$, there
exist two other solutions: $\phi_{min}=\pm\sqrt{-6m/g}$ that spontaneously
break the symmetry $\phi\rightarrow-\phi$. Anyway, for $g\neq0$
SUSY is classically preserved. 

Another possibility is $g=0$
and $\lambda\neq0$, in which case the model is super-renormalizable. If $2a\lambda\leq m^{2}$, the equation $S=0$
has two real solutions $\phi_{min}=-\frac{m}{\lambda}\pm(\frac{m^{2}}{\lambda^{2}}-\frac{2a}{\lambda})^{1/2}$
and SUSY is preserved. If instead, $2a\lambda>m^{2}$, the minimum
of $V^{(0)}(\phi)$ occurs for $\phi_{min}=-\frac{m}{\lambda}$ (solution
of $dV^{(0)}/d\phi=SS'=0$, for which $S=\frac{m^{2}}{2\lambda}-a\neq0$)
and implies in $V(\phi_{min})=\frac{1}{8\lambda^{2}}(2\lambda a-m^{2})^{2}>0$,
showing a spontaneous breakdown of SUSY at classical level. 
When only $a$ and $\lambda$ are non null, solutions: 
$\phi=\pm(2|a/\lambda|)^{1/2}$ exist for $a\lambda<0$ and do not exist for $a\lambda>0$, showing a breakdown of SUSY, 
with $V(\phi_{min}=0)=a^2/2$. 

\section{\label{sec:FeynmanRules}The effective potential}

There are several methods by which one can calculate loop corrections to the effective
potential in ordinary field theory. We will
employ the Jackiw's functional method \cite{key-20} whose extension to superspace
is straightforward. The recipe is: shift
the quantum superfield $\Phi$ by a classical superfield $\phi_{cl}$
and consider the action
\begin{equation}
{\cal \hat{S}}[\Phi,\phi_{cl}]\equiv{\cal S}[\Phi+\phi_{cl}]-{\cal S}[\phi_{cl}]-\int d^{5}z\Phi\left.\frac{\delta {\cal S}}{\delta\Phi}\right|_{\Phi=\phi_{cl}},
\end{equation} where $\phi_{cl}(\theta)=\sigma_{1}-\theta^{2}\sigma_{2}$,
with $\sigma_{1}=\left\langle \phi\right\rangle $ and $\sigma_{2}=\left\langle F\right\rangle $
being the constant vacuum expectation values of the scalar component
fields (the Lorentz invariance of the vacuum requires that $\left\langle \psi^{\alpha}\right\rangle =0$). The action $\hat {\cal S}$ takes the form
\begin{eqnarray}
{\cal \hat{S}}[\Phi,\phi_{cl}] & = &\int d^{5}z\left[\frac{1}{2}\Phi\left(D^{2}+m+\lambda\phi_{cl}+\frac{g}{2}\phi_{cl}^{2}\right)\Phi\right.\nonumber \\
 &+& \left.\frac{1}{3!}(\lambda+g\phi_{cl})\Phi^{3}+\frac{g}{4!}\Phi^{4}\right].\label{eq:4.2}
\end{eqnarray}

The effective potential can be written in a manifestly supercovariant
form as
\begin{eqnarray}
V_{eff}(\sigma_{1},\sigma_{2})&=&V^{(0)}(\sigma_{1},\sigma_{2})-\frac{i}{2\Omega}\ln Det\left[i\Delta_{F}^{-1}(z,z')\right]\nonumber\\
&&+\frac{i}{\Omega}\left\langle 0\left|T\exp i\int d^{5}z\hat{\mathcal{L}}_{int}(\Phi,\phi_{cl})\right|0\right\rangle .\label{eq:4-1}
\end{eqnarray}
The first term in Eq. (\ref{eq:4-1}) is the tree-level potential
as given in (\ref{eq:3}). The second term is the one-loop correction,
where 
\begin{eqnarray}
i\Delta_{F}^{-1}(z,z')&=&\left.\frac{\delta^{2}S[\Phi]}{\delta\Phi_{z}\delta\Phi_{z'}}\right|_{\Phi=\phi_{cl}}\nonumber\\
&=&\left(D^{2}+m+\lambda\phi_{cl}+\frac{g}{2}\phi_{cl}^{2}\right)\delta^{5}(z-z'),\label{eq:5-1}
\end{eqnarray}
and $\Omega\equiv\int d^{3}x$ is the spacetime volume. The
third term encodes the higher loops corrections: the sum of one-particle-irreducible vacuum superdiagrams with two and more loops computed from the shifted action (\ref{eq:4.2}). Let us note that the effective potential is
only function of the constant $ (x^{\mu}$ independent fields $\sigma_{1}$ and $\sigma_{2}$.
Actually, the superfield approach adopted here guarantees that after
all the $D$-algebra manipulations, only a single $\theta$-integration
remains to be done. This allows us to read off the effective potential
as it was previously made in (\ref{eq:2}) and (\ref{eq:3}).

\subsection{One-loop contribution}

The one-loop contribution $V^{(1)}$ to the effective potential is enclosed by the functional determinant in (\ref{eq:4-1}). It can be evaluated by the $\zeta$-function method as described in Ref. \cite{key-13}. Following the calculations outlined in Appendix (\ref{sec:Appendix-A:-TheZetaFuction})  we get
\begin{eqnarray}
V^{(1)} &=& -\frac{i}{2}\int\frac{d^{3}k}{(2\pi)^{3}}\ln\left[\frac{k^{2}+M^{2}}{k^{2}+\mu_{1}^{2}}\right]\nonumber\\
	&=& \frac{1}{12\pi}\left[ (\mu_{1}^2)^{3/2}-  (M^2)^{3/2}\right],
\label{eq:pot1}
\end{eqnarray} 
where dimensional reduction with minimal subtraction was used to perform 
the integrals. The parameter $\mu_{1}=S'$ is the fermionic mass, $\mu_{2}^{2}=\sigma_{2} S''$ (note that $\mu_{2}^{2}$ may assume positive or negative values) and $M^{2}=\mu_{1}^{2}-\mu_{2}^{2}=S'^2-\sigma_2 S''$ is the squared bosonic mass. It must also be noted that the perturbative calculation is valid only for $M^2$ positive (for $M^2<0$ the effective potential becomes complex). The prime denotes derivation with respect to $\sigma_1$. Therefore, up to one-loop order, the effective potential is given by
\begin{eqnarray}
V_{eff}(\sigma_{1},\sigma_{2})&=&-\frac{1}{2}\sigma_{2}^{2}-\sigma_{2}S\nonumber\\
&&+\frac{2}{3}\alpha\left[ (S'^2)^{3/2}-
(S'^2-\sigma_2 S'')^{3/2}\right]+ {\cal O}(\alpha^2)\,,\nonumber\\ 
\label{eq:Veff}
\end{eqnarray}
where we defined $\alpha=\hslash /8\pi=1/8\pi$ as the parameter that characterizes the strength of the one loop terms. The ${\cal O}(\alpha^2)$ stand for higher loops orders of approximation.

Let us now investigate the possibility of SUSY breaking and the stability of the effective potential. The stationary points of $V_{eff}$ are determined from the conditions 
\begin{eqnarray}
0&=&\frac{\partial V_{eff}}{\partial \sigma_{2}}=-\sigma_2-S+\alpha S''(S'^2-\sigma_2S'')^{1/2}+{\cal O}(\alpha^2)
\label{stationarySig2},\\
0&=&\frac{\partial V_{eff}}{\partial \sigma_{1}}=-\sigma_{2}S'+\alpha \left[ 2S'S'' (S'^2)^{1/2}\right.\nonumber\\
&&\left. -(2S'S''-g\sigma_2)
(S'^2-\sigma_2S'')^{1/2}\right]+{\cal O}(\alpha^2) \label{stationarySig1}.
\end{eqnarray}
As we are calculating the effective potential in loops approximations (powers of $\alpha$) we must, for consistency, solve 
(\ref{stationarySig2}) perturbatively, as a power series in $\alpha$ (see discussion below in this section and in Sec.V of \cite{key-137}). By substituting the trial form $\sigma_2=-S+\alpha A(\sigma_1)+{\cal O}(\alpha^2)$ in \eqref{stationarySig2}, we get:
\begin{equation}
\sigma_2(\sigma_1)=-S+\alpha \, S''(S'^2+S''S)^{1/2}+{\cal O}(\alpha ^2).\\
\label{eq:Sigma2}
\end{equation}

If SUSY is preserved the minimum of the effective potential must be $V_{eff}=0$, occurring for some real $\sigma_1$ and for  $\sigma_2=0$ 
(which means that the bosonic and fermionic masses, $M$ and $\mu_1$, remain equals). As can be seen, \eqref{eq:Veff} and \eqref{stationarySig1} are identically satisfied for $\sigma_2=0$. So, for SUSY to be preserved, \eqref{stationarySig2} must have a solution
$\sigma_2=0$, what means that the equation 
\begin{equation}
0=-S+\alpha S''(S'^2+S''S)^{1/2}+{\cal O} (\alpha ^2),\\
\label{eq:Sigma2a}
\end{equation}
must have a real solution  $\sigma_1=\bar \sigma_{1}$. In this case, the field configuration $(\sigma_{1}=\bar \sigma_{1},\sigma_{2}=0)$ is both a
stationary point and a zero of $V_{eff}$. 
If instead,  this equation does not have a real solution for $\sigma_1$, then $\sigma_2=0$ is not a solution 
of \eqref{stationarySig2} and SUSY is broken.

Suppose that $\bar \sigma_1$ does exist. Inserting the solution \eqref{eq:Sigma2} back in the effective potential, we get the ``physical" effective potential:
\begin{eqnarray}
U_{eff}(\sigma_1)&=&V_{eff}(\sigma_1,\sigma_2(\sigma_1))\nonumber\\ 
&=&\frac{1}{2} S^2 +\frac{2}{3} \alpha \left( (S'^2)^{3/2}-(S'^2+SS'')^{3/2}\right)+{\cal O} (\alpha^2).\nonumber\\
\label{Ueff}
\end{eqnarray}

It lacks yet, to determine if \eqref{eq:Sigma2} does have a solution $\bar \sigma_1$ and if $U_{eff}(\sigma_1)\geqslant 0$ in the region around 
$\sigma_1=\bar \sigma_{1}$, in which we can trust the loop calculation. 
As already observed, if the equation (\ref{eq:Sigma2a}) does not have a real solution for $\sigma_1$, then SUSY is broken.
So, let us start by analyzing the solutions of \eqref{eq:Sigma2a}. Moving $S$ to the left side, taking the square, and solving for $S$, we get: 
$S \mp \alpha S'S''={\cal O}(\alpha^2)$, for $S'=\pm|S'|$. Up to order $\alpha$ this equation reads:
\begin{eqnarray}
&&\left(a \mp\alpha m \lambda \right)+\left[ m \mp \alpha(gm+\lambda^2)\right] \sigma_1
\nonumber\\
&&\ \ +\frac{\lambda}{2}\left(1\mp 3\alpha g\right)\sigma_1^2+\frac{g}{6} \left( 1 \mp 3\alpha g \right)\sigma_1^3=0.
\end{eqnarray}

For $g\neq0$, this equation has at least one real solution for $\sigma_1$.
In the particular case $a=m=\lambda=0$ this (triple) solution is $\sigma_1=0$. If $a=\lambda=0$ we have a 
solution $\sigma_1=0$ and if additionally, $m/g <0$, two other solutions, the roots of $\sigma_1^2=-\frac{6m}{g}(1\mp2\alpha g)$,
which break the symmetry $\Phi \rightarrow -\Phi$, but not SUSY.

If $g=0$, real solutions exist if $m^2+\alpha^2 \lambda^4>2a\lambda$ and do not exist otherwise (dropping the term with $\alpha$ in this condition, we get back to the classical condition for SUSY preservation).

Many other particular cases can be studied, but we will fix in the more interesting case in which only the parameter $g\neq0$,
for which, the model is classically scale invariant. By substituting $S=\frac{g}{6}\sigma_1^3$ in \eqref{Ueff} we get
\begin{equation}
U_{eff}= \frac{g^2}{72}\sigma_1^6\left[1-\alpha \frac{g}{12}\left[\left(\frac{5}{3}\right)^{3/2}-1\right]\right]\,,\\
\end{equation}
which is positive (or null) for $g\ll 1$. So, for this subcase $(\sigma_1=0,\sigma_2=0)$ is the minimum of the effective potential and SUSY is preserved. 

For $g=m=0$ and $a\lambda<0$, Eq. \eqref{Ueff} becomes:
\begin{equation}
U_{eff}=\frac{1}{2}\left(a+\frac{\lambda}{2}\sigma_1^2\right)^2+\frac{2\alpha}{3}\left[\left(\lambda^2\sigma_1^2\right)^{3/2}-
\left(\frac{3}{2}\lambda^2\sigma_
1^2-|a\lambda|\right)^{3/2}\right].\\
\end{equation}

We must remember that the calculations can only be trusted for $m_B^2=\frac{3}{2}\lambda^2\sigma_1^2-|a\lambda|>0$, that is
for $\sigma_1^2>\frac{2}{3}|\frac{a}{\lambda}|$,
in which case $U_{eff}$ is positive (its zeros occur for $\sigma_1=\pm(2|\frac{a}{\lambda}|)^{1/2}\pm\alpha\lambda$).
In this case the discrete symmetry is broken and SUSY is preserved. For $a\lambda=|a\lambda|$ SUSY is broken, as in the classical case. 

These results are in accordance with that of Ref. \cite{key-15.2} where, using Wilson renormalization group equations, it is shown that 
SUSY is preserved for superpotentials with an even highest power of $\phi$ ($g\neq0$), but can or not be conserved (depending on 
the relation among the parameters) for odd highest power of $\phi$ ($g=0$ and $\lambda \neq 0$).

An observation is in order. In Ref. \cite{key-15.3} the authors observe that the ``physical" effective potential 
is positive for any value of $\sigma_1$, if the auxiliary field $\sigma_2$ is eliminated by exactly solving its equation of motion (equation (\ref{stationarySig2}), in the present paper). As they say, this positivity must result from effects of higher orders in $\alpha$, involved in the 
exact solution of (\ref{stationarySig2}). We did not try to confirm this claim; we instead, took the viewpoint that equation (\ref{stationarySig2}) is valid up to first order in $\alpha$ and so, its solution (our (\ref{eq:Sigma2})) must also be trusted up to this same order in $\alpha$ (an interesting discussion about these alternative views is given in Sec. 5 of Ref. \cite{key-137}). In the approximation that we are considering, the solution of  (\ref{stationarySig2}) can becomes complex (for values of $\sigma_1$ so that $SS''+S'^2<0$) and imply that the effective potential becomes complex 
in the region in which the classical potential ($U=S^2/2$) is not convex. This is a characteristic of loop calculations  and
not a particularity of SUSY \cite{key-15-4}.

\subsection{Two-loop contribution}

As is well known, for symmetry breaking to occurs by radiative corrections, we need the induction of terms of the form $h(\sigma_{1},\sigma_{2})\ln f(\sigma_{1},\sigma_{2})$.
In $2+1$ dimensions, this only happens in two (or more) loops approximation.
To study this possibility and to make a detailed analysis of the UV counterterms needed
to renormalize the effective potential, we will consider the general
case in which all the parameters in Eq. (\ref{eq1}) are non-null. 

Let us start by establishing the supergraph Feynman rules for the shifted
theory (\ref{eq:4.2}). The Feynman propagator satisfies the Green
equation:
\begin{equation}
\hat{\mathcal{O}}_{z}\Delta_{F}(z-z')=i\delta^{5}(z-z')\label{Operator},
\end{equation}
where $\hat{\mathcal{O}}_{z}=D_{z}^{2}+\mu_{1}-\mu_{2}^{2}\theta^{2}$
with $\mu_{1}$ and $\mu_{2}^{2}$ defined as before. 

To invert the operator $\hat{\mathcal{O}}$ we make use of the projection operators method, developed in Ref. \citep{key-21}. 
A basis for the space of scalar operators
is formed by the set of six linearly independent operators: 
$$
\begin{array}{lll}
P_{0}=1,  & P_{1}=D^{2},  & P_{2}=\theta^{2},\\
P_{3}=\theta^{\alpha}D_{\alpha}, & P_{4}=\theta^{2}D^{2}, & P_{5}=i\partial_{\alpha\beta}\theta^{\alpha}D^{\beta},\\
\end{array}
$$satisfying the multiplication table shown in Table \ref{tab:table1}.
\begin{center}
\begin{table}[!]
\begin{centering}
\begin{tabular}{|c|c|c|c|c|c|}
\hline 
 & $P_{1}$  & $P_{2}$  & $P_{3}$  & $P_{4}$  & $P_{5}$ \tabularnewline
\hline 
$P_{1}$  & $\square$  & $-P_{0}+P_{3}+P_{4}$  & $2P_{1}+P_{5}$  & $-P_{1}+\square P_{2}-P_{5}$  & $\square(-2P_{0}+P_{3})$ \tabularnewline
\hline 
$P_{2}$  & $P_{4}$  & $0$  & $0$  & $0$  & $0$ \tabularnewline
\hline 
$P_{3}$  & $-P_{5}$  & $2P_{2}$  & $P_{3}-2P_{4}$  & $2P_{4}$  & $2\square P_{2}+P_{5}$ \tabularnewline
\hline 
$P_{4}$  & $\square P_{2}$  & $-P_{2}$  & $2P_{4}$  & $-P_{4}$  & $-2\square P_{2}$ \tabularnewline
\hline 
$P_{5}$  & $-\square P_{3}$  & $0$  & $-2\square P_{2}+P_{5}$  & $0$  & $\square(P_{3}+2P_{4})$ \tabularnewline
\hline 
\end{tabular}
\end{centering}
\caption{\label{tab:table1} Multiplication table employed in the inversion of $\mathcal{\hat{O}}$.
In addition, one have the trivial relations: $P_{0}P_{i}=P_{i}P_{0}=P_{i}$,
with $i=0,\ldots,5$.}
\end{table}
\end{center}
After a straightforward algebra, the superpropagator in momentum space is given by
\begin{equation}
\Delta_{F}(k;\theta-\theta')=i\left(\sum_{i=0}^{5}c_{i} P_{i}\right)\delta^{2}(\theta-\theta'),
\end{equation}where:\begin{eqnarray*}
& & c_{0}=\frac{\mu_{1}}{k^{2}+M^{2}},\ \  c_{1}=-\frac{1}{k^{2}+M^{2}},\\
& & c_{2}=-\frac{(k^{2}-\mu_{1}^{2})\mu_{2}^{2}}{(k^{2}+\mu_{1}^{2})(k^{2}+M^{2})},\ \
c_{3}=-\frac{\mu_{1}\mu_{2}^{2}}{(k^{2}+\mu_{1}^{2})(k^{2}+M^{2})},\\
& & c_{4}=-\frac{2\mu_{1}\mu_{2}^{2}}{(k^{2}+\mu_{1}^{2})(k^{2}+M^{2})},\ \
c_{5}=-\frac{\mu_{2}^{2}}{(k^{2}+\mu_{1}^{2})(k^{2}+M^{2})}.
\end{eqnarray*}

The interaction vertices may be read from Eq. (\ref{eq:4.2}) and
the symmetry factors can be determined by the Wick's theorem in the
conventional way. 
\begin{center}
\begin{figure}[!]
\begin{centering}
\includegraphics[scale=0.5]{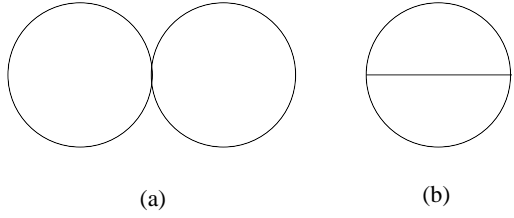}
\end{centering}
\caption{\label{fig:Two-loop-vacuum-bubbles}Two-loop vacuum bubble supergraphs.}
\end{figure}
\end{center}

The two-loop superdiagrams  contributing for the effective potential
are drawn in Fig. \ref{fig:Two-loop-vacuum-bubbles}. The associated
analytical expressions are shown in Appendix (\ref{sec:Appendix-B:-Two-loopDiagrams}) and the resulting two-loop momentum integrals are evaluated by dimensional
reduction using the formulas presented in Ref. \cite{key-22}.

The contribution of the diagram (a), denoted by $V_{a}^{(2)}$,
results to be finite, since it is constituted by the product of nonoverlapping
one-loop integrals. The diagram (b), instead has divergences proportional
to all the terms present in the tree-level potential $V^{(0)}$, which
is consistent with the usual renormalizability of the model. In summary,
we have the following results:
\begin{eqnarray}
V_{a}^{(2)} & = & -\frac{g}{32\pi^{2}}\frac{M\mu_{1}\mu_{2}^{2}}{(M+\mu_{1})},\nonumber\\
V_{b}^{(2)} & = & \frac{(\lambda+g\sigma_{1})^{2}}{64\pi^{2}}\left[\frac{\mu_{2}^{2}}{2}I_{div}-6\mu_{1}^{2}\ln\left(\frac{2M+\mu_{1}}{\mu}\right)+(M^{2}+5\mu_{1}^{2})\ln\left(\frac{3M}{\mu}\right)\right]\nonumber \\
 & + & \frac{(\lambda+g\sigma_{1})^{2}}{64\pi^{2}}\left[-M^{2}\ln\left(\frac{M}{\mu}\right)+\frac{M^{2}}{3}\left\{ 1+\ln\left(\frac{M+2\mu_{1}}{27\mu}\right)\right\} -\frac{2}{3}M\mu_{1}\right.\nonumber \\
 & + & \left.\frac{\mu_{1}^{2}}{3}\left\{ 1-6\ln\left(\frac{3M}{\mu}\right)-10\ln\left(\frac{M+2\mu_{1}}{\mu}\right)\right\} +\frac{2}{3}(M^{2}+8\mu_{1}^{2})\ln\left(\frac{2M+\mu_{1}}{\mu}\right)\right]\nonumber \\
 & + & \frac{g\sigma_{2}}{64\pi^{2}}\left[\left\{ I_{div}-2\ln\left(\frac{3M}{\mu}\right)\right\} \left(\lambda\mu_{1}+g\mu_{1}\sigma_{1}-\frac{g}{6}\sigma_{2}\right)\right],
\label{eq:TwoLoopdiv}
\end{eqnarray}
where $I_{div}=\frac{1}{\epsilon}+\ln[4\pi e^{(1-\gamma_{E})}]$ and
$\mu$ is an arbitrary mass parameter introduced via dimensional regularization.

The effective potential up to two loops is given by
\begin{equation}
V_{eff}=V^{(0)}+V^{(1)}+V_{a}^{(2)}+V_{b}^{(2)}+V_{CT},
\end{equation}
in which $V_{CT}$ is the counterterm contribution to the potential 
\begin{equation}
V_{CT}=-\left[\frac{1}{2}\delta Z\sigma_{2}^{2}+\delta m\,\sigma_{1}\sigma_{2}+\frac{\delta\,\lambda}{2}\sigma_{1}^{2}\sigma_{2}+g\frac{\delta g}{6}\sigma_{1}^{3}\sigma_{2}+\delta a\,\sigma_{2}\right],\label{eq:counterterms}
\end{equation}
as can be read from the classical Lagrangian in Eq.(\ref{eq:3}); $\delta Z$ is the wave function renormalization counterterm and the other counterterms are self explaining.

The divergent parts of $V^{(2)}$ can be collected in
\begin{eqnarray}
V_{div}^{(2)}& = & \frac{I_{div}}{128\pi^{2}}\left[-\frac{1}{3}g^{2}\sigma_{2}^{2}+(2g^{2}m+5g\lambda^{2})\sigma_{1}\sigma_{2}
\right.\nonumber\\
&& \left.  + 6g^{2}\lambda\,\sigma_{1}^{2}\sigma_{2}+2g^{3}\sigma_{1}^{3}
\sigma_{2}+(2gm\lambda+\lambda^3)\sigma_2 \right].\label{eq:divergentParts}
\end{eqnarray}

As seen from this equation, the renormalization of the effective potential requires all the counterterms in 
Eq. (\ref{eq:counterterms}):
\begin{eqnarray}
\delta Z &=&-\frac{1}{3}\hat{g}^2I_{div}+\delta Z_{fin}\nonumber\\
\delta a &=&\frac{1}{2}(2m\hat{g}\hat{\lambda}+\lambda \hat{\lambda}^2)I_{div}+\delta a_{fin}\nonumber\\
\delta m &=&\frac{1}{2}(2m \hat{g}^2+5g \hat{\lambda}^2)I_{div}+\delta m_{fin}\nonumber\\
\delta \lambda &=& 6\hat{g}^2 \lambda I_{div}+\delta \lambda_{fin}\nonumber\\
\delta g &=& 6\hat{g}^2I_{div}+\delta g_{fin},
\label{counterterm}
\end{eqnarray}
where we defined $\hat{g}=g/8\pi$ and $\hat{\lambda}=\lambda/8\pi$.

Let us compare our results with some others in the literature. In \cite{key-12.1} the effective potential of the
${\cal N} =2$ WZ model in 2+1 D was studied in two loops approximation. The authors 
conclude that only a wave function renormalization is needed.  
As they say, that result is not unexpected; the ${\cal N}=2$ superspace formulation of supersymmetry in 2+1 D 
can be got from the ${\cal N}=1$ superspace formulation in 3+1 D by dimensional reduction and so, 
the 3+1 D nonrenormalization theorems are expected to work with ${\cal N}=2$ in 2+1 D supersymmetry.
In our results on the other side, no nonrenormalization theorem applies and the renormalization of all the 
parameters are necessary. Differently from ours, in which three different arguments appear in the generated logarithms, 
their expression has only a single argument in the generated logarithms. This difference is, maybe, due 
to their approximation, in which spinorial derivatives $D_{\alpha}\Phi$ and $D^2 \Phi$, besides
the usual spatial $\partial \Phi/\partial x^{\mu}$ are dropped during the calculations.
Our results also contradict the result for a similar ${\cal N}=1$ model, reported in Ref. \cite{key-14}, in
which a counterterm of the form $\sigma_{1}^{6}$, not present in the classical Lagrangian, was found to be required.

For the model with $g\neq0$ and $\lambda\neq0$ the
renormalization also requires that $\delta a$ and $\delta m$ be
non null. The sub model with only $g\neq0$  is renormalizable, that is, it only requires the 
renormalization of $g$ besides that of $Z$. We will study this subcase in the next section.

If $g=0$ and $\lambda\neq0$ (in which case the model is
super-renormalizable) the cancellation of the UV divergences, up two
loops, only requires  that $\delta a\neq0$ ($\delta a=\frac{1}{2}\lambda {\hat \lambda}^2 I_{div}$).
As the divergent parts of $\delta \lambda$, $\delta m$ and $\delta Z$ are zero, no running of these 
constants or an anomalous scaling of the field occur; these results disagree with those in reference \cite{key-15.2}.
This fact is not surprising, considering that the involved approximations in the two methods of calculation are very different. 
In the two loops approximation, the only parameter that runs with the scale is $a$. The renormalization group equation for $a$ is obtained from the relation between the unrenormalized $a_0$ and the 
renormalized $a$, which is given by:
\begin{eqnarray}
a_0&=&\mu^{-\epsilon /2} \frac{a+\delta a}{(1+\delta Z)^{1/2}}\nonumber\\
&=&\mu^{-\epsilon /2}\left[a+\ln(4\pi e^{1-\gamma})+\frac{1}{2}\lambda 
{\hat \lambda}^{2} \frac{1}{\epsilon}+\cdots \right].
\end{eqnarray}
From the equation $0=\mu(\partial a_0/\partial \mu)$, we get $\mu(\partial a/\partial \mu) =\lambda {\hat \lambda}^{2}/4$, which after integration gives
\begin{equation} 
a(\mu)=a(\mu_0)+\frac{\lambda {\hat \lambda}^{2}}{4} \ln\left(\frac{\mu}{\mu_0}\right).
\end{equation}
This result means that a change in the parameter $\mu$ can be compensated by a simultaneous change in $a$, leaving the effective
potential invariant.

\section{\label{sec:4}The unbroken susy vacuum}

Let us now  investigate in more details the sub model 
with $g\neq0$ and $m=\lambda=a=0$, which is of particular interest for being classically scale invariant. 
As discussed in the previous section, the model only requires the $\delta Z$ and $\delta g$ counterterms.
The total renormalized effective potential $V_{eff}$ takes the form
\begin{eqnarray}
V_{eff}&=&-\frac{1+\delta Z_{fin}}{2} \sigma_2^2- g\frac{1+\delta g_{fin}}{6} \sigma_1^3 \sigma_2+\frac{2\alpha}{3}(\mu_1^3-M^3)\nonumber\\
&+&2\alpha^2 g\left[\frac{1}{3}\mu_1(\mu_1-M)(\mu_1-4M)
-\frac{2}{3}\mu_1(\mu_1^2-M^2)\ln\left(\frac{2M+\mu_1}{\mu}\right)\right.\nonumber\\
 &-&\left.\frac{1}{3}\mu_1(10\mu_1^2-M^2)\ln\left(\frac{M+2\mu_1}{\mu}\right)+\left(\mu_1(2\mu_1^2+M^2)+\frac{g}{6}\sigma_2^2\right)\ln\left(\frac{3M}{\mu}\right)\right],
\end{eqnarray}
where $\mu_1=g\sigma_1^2/2$, $\mu_2^2=g\sigma_1\sigma_2$ and $M=(\mu_1^2-\mu_2^2)^{1/2}$. The parameters $\alpha$ and $\alpha^2$ indicate the contributions of one and two loops. Observe that $V_{eff}$ is 
real only for $M$ real, that is, if $(g\sigma_1^4-4\sigma_1\sigma_2)>0$. The singularity in $\sigma_1=0$, for 
$\sigma_2\neq0$, in the last term of $V_{eff}$, is a reminiscence of the IR divergences due to the null mass of the model. So, $\sigma_1=0$ is not a convenient spot to impose renormalization conditions. The point $\sigma_1^2=\mu$, where $\mu$ is the mass parameter introduced by the dimensional regularization, is a more natural spot. To see this fact, let us expand the expression of the effective potential in powers of $\sigma_2$. The result is
\begin{eqnarray}
V_{eff}&=&-\frac{g}{6}\sigma_{2}\sigma_{1}^{3}\left[1+\left(\delta g_{fin}-3\hat{g}+9\hat{g}^{2}+12\hat{g}^{2}\ln\left(\frac{3g}{2}\right)\right)+12\hat{g}^{2}\ln\left(\frac{\sigma_{1}^{2}}{\mu}\right)\right]\nonumber\\
&-&\frac{1}{2}\sigma_{2}^{2}\left[1+\left(\delta Z_{fin}+\hat{g}-\frac{29}{9}\hat{g}^{2}-\frac{2}{3}\hat{g}^{2}\ln\left(\frac{3g}{2}\right)\right)-\frac{2}{3}\hat{g}^{2}\ln\left(\frac{\sigma_{1}^{2}}{\mu}\right)\right]+\sigma_{2}^3\,\mathcal{F}(\sigma_{1},\sigma_{2}),
\end{eqnarray}
where, as before, $\hat g =g/8\pi$. We choose $\delta g_{fin}$ and $\delta Z _{fin}$ by imposing that the terms in the parentheses be nulls. These choices imply that at the point $\sigma_1^2=\mu$, the coefficients 
of the two monomials ($\sigma_2\sigma_1^3$ and  $\sigma_2^2$) are 
the same as in the classical potential $V_{cl}=-(g/6)\sigma_2\sigma_1^3-(1/2)\sigma_2^2$. The first condition fix the renormalized coupling constant and the second implies 
that the coefficient of the kinetic term of the effective renormalized Lagrangian at $\sigma_1=\mu$ is one.
In the expanded form, the renormalized potential results in
\begin{eqnarray}
V_{eff}&=&-\frac{g}{6}\sigma_2 \sigma_1^3\left(1+12\hat g^2 \ln\left(\frac{\sigma_1^2}{\mu}\right)\right)\nonumber\\
&-&\frac{1}{2}\sigma_2^2 \left(1-\frac{2}{3}\hat g^2 \ln\left(\frac{\sigma_1^2}{\mu}\right)\right)+
\sigma_2^3\,{\cal F} (\sigma_1,\sigma_2)\,.
\label{eq:Veff3}
\end{eqnarray}
 In the previous section we analyzed the effective potential up to one-loop order with minimal subtractions ($\delta Z_{fin}=\delta g_{fin}=0$). In the present section we made finite renormalizations, so that in the expansion up to the second power of 
$\sigma_2$, no one loop correction survived; the only corrections to the classical potential come from the two-loops order.

Let us now investigate the possibility of supersymmetry breakdown. It is
easy to check that $V_{eff}(\sigma_{1},\sigma_{2}=0)=0$,
from which it also follows that $\left.\partial V_{eff}/\partial\sigma_{1}\right|_{\sigma_{2}=0}\equiv0$. 
The condition $\left.\partial V_{eff}/\partial\sigma_{2}\right|_{\sigma_{2}=0}=0$,
leads to the following (gap) equation for $\sigma_{1}$:
\begin{equation}
\sigma_1^3\left[1+12\hat g^2 \ln\left(\frac{\sigma_1^2}{\mu}\right)\right]=0.
\end{equation}
This equation has a trivial solution $\sigma_1^{min}=0$ that ensures
that SUSY as well as the
discrete symmetry are not broken by the radiative corrections. Looking
at the term in the parentheses, a possible non-zero solution $\sigma_{1}^{min}\neq0$ would be given by
\begin{equation}
1+12\hat g^2 \ln\left(\frac{\sigma_{1}^{2}}{\mu}\right)=0.
\end{equation}
However, by looking at (\ref{eq:Veff3}) we see that the two loops corrections are proportional to  $\hat g^2 \ln\left(\sigma_1^{2}/\mu\right)$  
which, for the validity of the perturbative approach, must be small as compared to the factor (one) coming from the zero loops potential.
So, this minimum lies very far from the range of validity of the two loops approximation; we conclude that,
no non-trivial vacuum is induced by radiative corrections and no SUSY breaking nor mass generation occur. 
This result contradicts the claim made in Ref. \cite{key-15}, that 
the two-loop corrections are able to induce supersymmetry breaking and dynamical generation
of mass. On the other hand a similar conclusion to ours was obtained in \cite{key-15.1}, for the O(N) WZ model in $1/N$ 
approximation. The same conclusion was also got in \cite{key-15.2} through a functional renormalization group analysis.  
In fact, as discussed in the seminal paper \citep{key-12}, by S. Coleman and E. Weinberg, 
spontaneous symmetry breaking and mass generation, through radiative corrections, can only occur 
in models with more than one coupling constant and is made possible through an interplay among these constants.

In two loops, the equation $0=\partial V/\partial \sigma_2$ is a transcendental equation. Yet, a solution as a power series in $\alpha$ 
can be obtained and inserted back in $V$ to get the physical potential up to order $\alpha^2$. The solution for $\sigma_2$ is of the form
$\sigma_2=  C_1 \sigma_1^3+C_2\sigma_1^3 \ln (\sigma_1^2/\mu)+{\cal O}(\alpha^3)$, where $C_1$  and $C_2$ are functions of  
$\alpha$, $g$, $\delta Z$ and $\delta g$. The potential results in the form 
$U_{eff}=c_1 \sigma_1^6 +c_2 \sigma_1^6 \ln(\sigma_1^2/\mu)+{\cal O}(\alpha^3)$ 
with $c_1$ and $c_2$ to be fixed by renormalization conditions. The detailed analysis does not give any new information in relation to our previous and simpler discussion.}

Finally, let us determine the renormalization group function $\beta_{g}$ for the particular case with $g\neq0$ and $m=\lambda=a=0$. 
Introducing the bare $\Phi_0$ and renormalized superfield $\Phi$ and the renormalized
coupling constant $g$ through the definitions
\begin{eqnarray}
\Phi_{0} & = & Z_{\Phi}^{\frac{1}{2}}\Phi=(1+\delta Z)^{\frac{1}{2}}\Phi,\label{eq:deltaZ}\\
g_{0} & = & \mu^{\varepsilon}gZ_{g}=\mu^{\varepsilon}g\left[\frac{1+\delta g}{Z_{\Phi}^{2}}\right],\label{eq:deltag}
\end{eqnarray}
and writing explicitly the counterterms from Eq. (\ref{counterterm})
as 
\begin{eqnarray}
\delta Z & = & -\frac{g^{2}}{192\pi^{2}}\frac{1}{\varepsilon}+\mbox{finite},\label{eq:divdeltaZ}\\
\delta g & = & \frac{3g^{2}}{32\pi^{2}}\frac{1}{\varepsilon}+\mbox{finite},\label{eq:divdeltag}
\end{eqnarray}
we obtain the beta function at leading order
\begin{eqnarray}
\beta_{g} & = & \mu\frac{\partial g}{\partial\mu}=\frac{5g^{3}}{24\pi^{2}}-\varepsilon g\nonumber \\
  & = & \frac{5g^{3}}{24\pi^{2}}\ \ \ \ \ \ \ (\mbox{for }\varepsilon\rightarrow0).\label{eq:betafunction}
\end{eqnarray}
This result is in agreement with that obtained in Ref. \cite{key-17} by calculating the divergent parts of several
vertex functions in the component fields formalism.
The solution of Eq. (\ref{eq:betafunction}) is given by
\begin{equation}
\bar{g}^{2}=\frac{g^{2}}{1-\frac{5}{12\pi^{2}}g^{2}\ln\left(\frac{\bar{\mu}}{\mu}\right)}.
\end{equation}
Starting with a $g^{2}\ll1$ at a scale $\mu$, we see that the effective
coupling constant $\bar{g}^{2}$ increases as the scale $\bar{\mu}$
is increased showing a Landau pole at some scale $\bar{\mu}$. So, at short distances the above results are not reliable:
higher loop corrections become more and more important compared to
the second order. If instead, we make $\bar{\mu}\rightarrow0$ we
get $\bar{g}^{2}\rightarrow0$, showing an IR free limit.

An anomalous scaling of the model is also induced as can be seen by calculating the anomalous dimension of the field:\begin{equation}
\gamma_{\Phi}=\frac{1}{2}\mu\frac{d\ln Z_{\Phi}}{d\mu}
\label{eq:gammafunction}.
\end{equation}

From (\ref{eq:deltaZ}), we can write (\ref{eq:gammafunction}) in the form
\begin{equation}
2\left(1+\delta Z\right)\gamma_{\Phi}=\mu\frac{\partial\delta Z}{\partial g}\frac{\partial g}{\partial\mu}.
\label{eq:gamma2}
\end{equation}
By replacing (\ref{eq:divdeltaZ}) and (\ref{eq:betafunction}) into (\ref{eq:gamma2}) we get
\begin{equation}
2\left(1-\frac{g^{2}}{192\pi^{2}}\frac{1}{\varepsilon}\right)\gamma_{\Phi}=\frac{g^{2}}{96\pi^{2}}-\frac{5g^{4}}{24\cdot 96\pi^{4}}\frac{1}{\varepsilon},
\end{equation} 
which yields $\gamma_{\Phi}=\frac{g^{2}}{192\pi^{2}}$.

\section{\label{sec:conclusions}Conclusions}

In the present paper, we calculate the effective potential for the ${\cal N}=1$
WZ model in $2+1$ dimensions. We employ
the Jackiw's functional method combined with the superfields formalism. A detailed analysis of the renormalizability and
vacuum structure of the model is presented, up to two loops. One of the main results is that the
renormalization of the theory requires, besides the wave function counterterm, also mass and
coupling constants counterterms, but not any new one. This result differs from that reported in \cite{key-14},
where the renormalization of the model requires an extra $\sigma_{1}^{6}$ counterterm.
It also differs from that in Ref. \cite{key-12.1} for the ${\cal N}=2$ WZ model in $2+1$ D in which only a 
wave function renormalization was found to be required.
For the massless $\Phi^{4}$ (sub) model, we also determined the $\beta_{g}$
function which agrees with the results of
\cite{key-17}, showing a Landau pole in the UV limit. At the same time, we found that the quantum
vacuum state preserves supersymmetry and the discrete symmetry $\Phi\rightarrow-\Phi$ of the classical
theory, contrary to the remark in \cite{key-15}, but in agreement with the results in Refs. \cite{key-15.1} and \cite{key-15.2}. 
A group renormalization study of the pure $g\neq0$ model, besides the calculation of the effective potential for the 
${\cal N}=2$ model will be addressed in a forthcoming paper.

\section*{ACKNOWLEDGMENTS}

This work was partially supported by the Brazilian agency Conselho Nacional de Desenvolvimento
Cient\' ifico e Tecnol\' ogico (CNPq) and by CAPES-Brazil. The authors thank E. A. Gallegos for reading the manuscript and useful suggestions.

\appendix
\section{\label{sec:Appendix-A:-TheZetaFuction}The $\zeta$-function
method}

In this appendix we compute the one-loop contribution $V^{(1)}$ by
the $\zeta$-function method following Ref. \cite{key-13}.
The functional determinant $Det\hat{\mathcal{O}}$ is understood as
the product of the eigenvalues of $\hat{\mathcal{O}}$. Starting with
the eigenvalues equation
\begin{equation}
\int d^{5}z'\hat{\mathcal{O}_{z}}(z,z')f_{n}(z')=\alpha_{n}f_{n}(z),
\end{equation}
and defining the $\zeta$-function associated to $\hat{\mathcal{O}}(z,z')\equiv\hat{\mathcal{O}}_{z}\delta^{5}(z-z')$
as
\begin{equation}
\zeta(s)=\sum_{n}\frac{1}{\alpha_{n}^{s}},
\end{equation}
the functional determinant of $\hat{\mathcal{O}}_{z}$ can be written
in the form 
\begin{equation}
Det\hat{\mathcal{O}}_{z}\equiv\prod_{n}\alpha_{n}=\exp\left[-\zeta^{\prime}(0)\right].
\end{equation}
So, the calculation of the determinant requires to get an analytic
representation for $\zeta(s)$ . To this end, let us introduce a two-point
superspace function $G(z,z';\tau)$ which obeys the equation:
\begin{equation}
\hat{O}_{z}G(z,z';\tau)+\frac{\partial G}{\partial\tau}=0,\label{eq:6-1}
\end{equation}
with the initial condition $G(x,\theta;x',\theta';\tau=0)=\delta^{3}(x-x')\delta^{2}(\theta-\theta').$ 

It is straightforward to check that

\begin{equation}
\zeta(s)=\frac{1}{\Gamma(s)}\int_{0}^{\infty}d\tau\tau^{s-1}\int d^{3}xd^{2}\theta G(x=x',\theta=\theta';\tau),\label{eq:4-2}
\end{equation}
for $G(z,z';\tau)\equiv\sum_{n}\exp[-\alpha_{n}\tau]f_{n}(z)f_{n}^{*}(z')$. 

To proceed we must now determine an explicit solution of $G(z,z';\tau)$
satisfying the Eq. (\ref{eq:6-1}) subject to the initial condition
above. To this aim, we will assume that this function is spacetime
translational invariant so that it can be written as:
\begin{equation}
G(x,\theta;x',\theta';\tau)=\int\frac{d^{3}k}{(2\pi)^{3}}g(k,\theta,\theta';\tau)\exp\left[-ik(x-x')\right],\label{eq:8}
\end{equation}
with the following ansatz for $g(k,\theta,\theta';\tau)$:
\begin{eqnarray}
g(k,\theta,\theta';\tau)&=& A(k,\tau)+\theta^{\alpha}\theta'^{\beta}k_{\alpha\beta}B(k,\tau)+\theta^{\alpha}\theta'_{\alpha}C(k,\tau)\nonumber\\&&+\theta^{2}D(k,\tau)+\theta'^{2}E(k,\tau)+\theta^{2}\theta'^{2}H(k,\tau).\nonumber\\
\label{eq:8-1}
\end{eqnarray}

To find the coefficients $A$, $B$, $C$, $D,$ $E$ and $H$, we
have to use the explicit form of $\hat{O}_{z}$ read off from Eq. \eqref{Operator}
and insert Eq. \eqref{eq:8-1} into Eq. \eqref{eq:6-1}. This equation splits
into six linear ordinary differential equations with the initial conditions:\begin{eqnarray}
A(k,0)=0 & B(k,0)=0 & C(k,0)=1\nonumber \\
D(k,0)=-1 & E(k,0)=-1 & H(k,0)=0,
\end{eqnarray}so that the solution of this system is readily found. From these results,
we now construct the $\zeta$-function as prescribed in Eq. (\ref{eq:4.2}).

After integration and using the relation $V^{(1)}=-(i/2\Omega)\ln Det\hat{\mathcal{O}} =(i/2\Omega)\zeta^{\prime}(0)$,
we are able to get the result described in Eq. \eqref{eq:pot1}.

\section{\label{sec:Appendix-B:-Two-loopDiagrams}Two-loop diagrams}

The analytical expressions for the two-loop vacuum bubbles that contribute
to the effective potential displayed in Fig. \ref{fig:Two-loop-vacuum-bubbles}
are $(d^{D}k\equiv\mu^{\varepsilon}d^{3-\varepsilon}k)$:
\begin{eqnarray}
V_{a}^{(2)} & = & -\frac{g}{8}\int\frac{d^{D}kd^{D}q}{(2\pi)^{2D}}d^{2}\theta\left.\Delta_{F}(k;\theta-\theta_{1})\right|_{\theta=\theta_{1}}\left.\Delta_{F}(q;\theta-\theta_{2})\right|_{\theta=\theta_{2}}\nonumber \\
 & = & -\frac{g}{2}\int\frac{d^{D}kd^{D}q}{(2\pi)^{2D}}\left[\frac{\mu_{1}\mu_{2}^{2}}{(k^{2}+M^{2})(q^{2}+\mu_{1}^{2})(q^{2}+M^{2})}\right],
\end{eqnarray}
and
\begin{eqnarray}
V_{b}^{(2)}&=&-3i\int\frac{d^{D}kd^{D}q}{(2\pi)^{2D}}d^{2}\theta_{1}d^{2}\theta_{2}\mathcal{I}(\theta_{1}^{2},\theta_{2}^{2})\Delta_{F}(k;\theta_{1}-\theta_{2})\nonumber\\
&&\times\Delta_{F}(q;\theta_{1}-\theta_{2})\Delta_{F}(-k-q;\theta_{1}-\theta_{2}),
\end{eqnarray}
where
\begin{eqnarray}
\mathcal{I}(\theta_{1}^{2},\theta_{2}^{2})&=&\frac{1}{36}\left[(\lambda+g\sigma_{1})^{2}-(\lambda g\sigma_{2}+g^{2}\sigma_{1}\sigma_{2})(\theta_{1}^{2}+\theta_{2}^{2})\nonumber\right.\\
&&\left.+g^{2}\sigma_{2}^{2}\theta_{1}^{2}\theta_{2}^{2}\right].
\end{eqnarray}

After performing the $D$-algebra and carrying out the remaining $\theta$-integration, we obtain the following two-loop momentum integrals
\begin{eqnarray}
V_{b}^{(2)} & = & \int\frac{d^{D}kd^{D}q}{(2\pi)^{2D}}\frac{-\mu_{2}^{2}(\lambda+g\sigma_{1})^{2}}{12(k{}^{2}+M^{2})(q{}^{2}+M^{2})(k{}^{2}+\mu_{1}^{2})(q{}^{2}+\mu_{1}^{2})\left[(k+q)^{2}+M^{2}\right]\left[(k+q)^{2}+\mu_{1}^{2}\right]}\nonumber \\
 && \times \left\{ k^{4}(q^{2}+\mu_{1}^{2})+2k.q\left[(k^{2}+\mu_{1}^{2})(q^{2}+\mu_{1}^{2})-(k^{2}+q^{2}-(k+q)^{2}+\mu_{1}^{2})\mu_{2}^{2}\right]\right.\nonumber \\
 &  & \ \ \  +\mu_{1}^{2}\left[q^{4}-15\mu_{1}^{4}-4q^{2}\left[(k+q)^{2}+2\mu_{1}^{2}\right]+6\mu_{1}^{2}\mu_{2}^{2}+2(k+q)^{2}(-5\mu_{1}^{2}+\mu_{2}^{2})\right]\nonumber \\
 &  & \ \ \ \left. + k^{2}q^{4}-4k^{2}\mu_{1}^{2}\left[(k+q)^{2}+2\mu_{1}^{2}\right]+k^{2}q^{2}\left[2(k+q)^{2}-\mu_{1}^{2}-4\mu_{2}^{2}\right]\right\} \nonumber \\
 & & + \int\frac{d^{D}kd^{D}q}{(2\pi)^{2D}}\frac{-6g\mu_{1}(\lambda+g\sigma_{1})\sigma_{2}+g^{2}\sigma_{2}^{2}}{12(k{}^{2}+M^{2})(q{}^{2}+M^{2})\left[(k+q)^{2}+M^{2}\right]}.
\end{eqnarray}

The two-loop integrals were performed by dimensional reduction
scheme, using the formulas from \cite{key-22}. The final results are written in Eq. \eqref{eq:TwoLoopdiv}.


\begin{thebibliography}{20}

\bibitem{key-1} S. Dimopoulos and H. Georgi, Nucl. Phys. B \textbf{193},
150 (1981); S. Dimopoulos, S. Raby and Frank Wilczek,  Phys. Rev.
D \textbf{24}, 1681 (1981). 

\bibitem{key-2} I. Affleck, M. Dine and N. Seiberg, Phys. Rev. Lett.
\textbf{51}, 1026 (1983); I. Affleck, M. Dine and N. Seiberg, Nucl.
Phys. B \textbf{241}, 493 (1984); I. Affleck, M. Dine and N. Seiberg,
N. Nucl. Phys. B \textbf{256}, 557 (1985).

\bibitem{key-3} Ann E. Nelson and Nathan Seiberg, Nucl. Phys. B \textbf{416}, 46 (1994).

\bibitem{key-4} E. Witten, Nucl. Phys. B \textbf{202}, 253 (1982).

\bibitem{key-5} A. V. Smilga, J. High Energy Phys. \textbf{01}, 086 (2010). 

\bibitem{key-6} L. O'Raifeartaigh, Nucl. Phys. B \textbf{96}, 331
(1975); P. Fayet and J. Illiopoulos, Phys. Lett. B \textbf{51}, 461
(1974).


\bibitem{key-8} S. Ray, Phys. Lett. B \textbf{642}, 137 (2006); K. Intriligator, N. Seiberg and D. Shih, J. High Energy Phys. \textbf{07}, 017 (2007).

\bibitem{key-10} M. Grisaru, M. Rocek and W. Siegel, Nucl. Phys.
B \textbf{159}, 429 (1979).

\bibitem{key-11} L. Alvarez-Gaumé, D. Z. Freedman and M. T. Grisaru, "Spontaneous Breakdown of Supersymmetry in two Dimensions,"
HUTMP 81/B111.

\bibitem{key-11.2} A. F. Ferrari, E. A. Gallegos, M. Gomes, A. C. Lehum, J. R. Nascimento, A. Yu. Petrov and A. J. da Silva, Phys. Rev. D \textbf{82}, 025002 (2010). 

\bibitem{key-12} S. Coleman and E. Weinberg, Phys. Rev. D \textbf{7}, 1888 (1973).

\bibitem{key-12.1} I.L. Buchbinder, B.S. Merzlikin, I.B. Samsonov, Nucl. Phys. B \textbf{860}, 87 (2012).

\bibitem{key-13} C. P. Burgess. Nucl. Phys. B \textbf{216} , 459 (1983).

\bibitem{key-14} D.G.C. McKeon and K. Nguyen, Phys. Rev. D \textbf{60},
085009 (1999).

\bibitem{key-15} A.C. Lehum, Phys. Rev. D. \textbf{77}, 067701 (2008).

\bibitem{key-15.1} A.C. Lehum, Phys. Rev. D. \textbf{84}, 107701 (2011).

\bibitem{key-15.3} F.  Synatschke, H. Gies and A. Wipf, Phys. Rev. D \textbf {80} 085007 (2009). 

\bibitem{key-15.2} F. Synatschke, J. Braun and A. Wipf, Phys. Rev. D \textbf {81} 125001 (2010). 

\bibitem{key-137} T. Murphy and L. O'Raifeartaigh, Nucl. Phys. B \textbf{218}, 484 (1983). 

\bibitem{key-15-4} Y. Fujimoto, L. O'Raifeartaigh and a G. Parravicini, Nucl. phys. B \textbf{212}, 268 (1983).

\bibitem{key-16} G. Fogleman and K. Viswanathan, Phys. Rev. D \textbf{30},
1364 (1984).

\bibitem{key-17} F. A. Dilkes, D. G. C. McKeon and K. Nguyen, Phys. Rev. D \textbf{57}, 1159 (1998).

\bibitem{key-18} S. J. Gates, M. T. Grisaru, M. Rocek and W. Siegel, \textit{Superspace, or One Thousand and One Lessons in Supersymmetry}, Frontiers in Physics, Vol. 58, (W. A. Benjamin, New York, 1983).

\bibitem{key-19} M. Drees, R. Godbole and P. Roy, \textit{Theory and
Phenomenology of Sparticles}, World Scientific Publishing, (Singapore,
2004).

\bibitem{key-20} R. Jackiw, Phys. Rev. D \textbf{9}, 1686 (1974).

\bibitem{key-21} J. L. Boldo, L. P. Colatto, M. A. De Andrade, O. M. Del Cima and J. A. Helay\"el-Neto, Phys. Lett. B \textbf{468},
96 (1999); E. A. Gallegos and A. J. da Silva, 	Phys. Rev. D \textbf{84}, 065009 (2011).

\bibitem{key-22} P. N. Tan, B. Tekin, and Y. Hosotani, Nucl. Phys. B \textbf{502}, 483 (1997); V. S. Alves, M. Gomes, S. L. V. Pinheiro and A. J. da Silva, Phys. Rev. D \textbf{61}, 065003 (2000); A. G. Dias, M. Gomes, and A. J. da Silva, Phys. Rev. D \textbf{69}, 065011 (2004).  


\end{thebibliography}
\end{document}